%% file: main.tex
\documentclass[twocolumn,superscriptaddress,groupedaddress,amsmath,amssymb,aps,pre,floatfix]{revtex4-2}

\usepackage{comment}
\usepackage{graphicx}
\usepackage{dcolumn}
\usepackage{bm}

\usepackage[utf8]{inputenc}
\usepackage[T1]{fontenc}

\usepackage{braket}
\usepackage{siunitx}
\usepackage{xcolor}

\usepackage[normalem]{ulem}

\begin{document}

\title{Extended Self-similarity in Multimode Optical Fiber Speckles}
\author{Mengxin Wu}
\author{Ziye Chen}
\author{Guang Yang}
\author{Mingshu Zhao}
\email{zmshum@ncepu.edu.cn}
\affiliation{Hebei Key Laboratory of Physics and Energy Technology, Department of Mathematics and Physics, North China Electric Power University, Baoding, Hebei 071003, China}

\date{\today} 

\begin{abstract}
Extended Self-Similarity (ESS) is a widely used tool for uncovering universal power-law scaling in systems dominated by nonlinear interactions. This work demonstrates that ESS scaling can also emerge in a system governed by purely linear physics: the propagation of coherent light in a multimode fiber. The system produces complex speckle patterns arising solely from deterministic linear mode interference. We analyze the intensity structure functions of these speckles and observe a robust extended scaling range. The measured scaling exponents align with the classical Kolmogorov scaling exponents. This finding establishes that the statistical signatures captured by ESS are not exclusive to nonlinear systems, revealing a broader applicability of this scaling framework to complex linear systems.
\end{abstract}

\maketitle

\section{Introduction}

Extended Self-Similarity (ESS) is a robust empirical scaling relationship that has profoundly advanced the analysis of turbulent and other complex, nonlinear systems~\cite{meneveau2000scale,alexakis2018cascades,marino2023scaling,jiang2019multifractal}.
Initially discovered in the context of classical fluid turbulence~\cite{benzi1993extended,benzi1993extended2,benzi1991multifractality}, it has since been established as a critical diagnostic tool in a range of physical systems dominated by nonlinearity and multi-scale interactions. 
Its validity and utility have been demonstrated in the complex dynamics of geophysical flows~\cite{nikora2001extended,carbone1996evidences,kiliyanpilakkil2016extended,zhao2025turbulence}, the superfluid turbulence of quantum fluids~\cite{zhao2025spatiotemporal,zhao2025kolmogorov,varga2018intermittency}, and the collective behavior of magnetized plasmas~\cite{budaev2008extended}. 
The principle of ESS has also been applied to non-physical systems, including the analysis of scaling in financial market fluctuations~\cite{constantin2005volatility}.
This recurrence across such diverse systems establishes ESS as a powerful and universal descriptor for scaling in complex systems.

The ESS technique was developed to address the limitations of directly observing power-law scaling in real turbulent flows.
In a fully developed turbulent flow, Kolmogorov's 1941 (K41) theory predicts an "inertial range" of scales where velocity structure functions exhibit universal power-law scaling~\cite{kolmogorov1941local}. However, in real-world systems with finite Reynolds numbers and strong intermittency, this ideal inertial range is often truncated or ill-defined. ESS circumvents this limitation by analyzing the relative scaling of structure functions of different orders against each other, rather than directly against the spatial scale~\cite{benzi1993extended}. 
This reveals a robust, extended scaling range, which facilitates a more reliable extraction of the scaling exponents that characterize the underlying intermittency.

While the ESS technique is empirically applied to turbulent systems—which are intrinsically nonlinear—the prevailing theoretical frameworks do not attribute its effectiveness directly to nonlinearity itself. Instead, these models explain its robust scaling through various proposed mechanisms, such as hierarchical structure or the role of the pressure-gradient term under mean-field approximation~\cite{yakhot2001mean, ching2002extended, chakraborty2010extended}.
This naturally raises a fundamental question: is nonlinearity a necessary prerequisite for ESS, or can similar scaling behavior emerge from purely linear interference? 
A test case for this question is presented by linear optical propagation in multimode fibers (MMF), a system where complex speckle patterns emerge from the linear interference of numerous guided modes~\cite{hill1980modal,daino1980speckle,rawson1980frequency,tremblay1981modal,takai1985statistical}. Determining whether ESS governs the statistics of these speckles would test the current understanding of its origins and potentially extend the concept of universal scaling to a new class of linear, disordered systems.

An MMF is a linear optical waveguide that supports a discrete yet large set of propagating modes. At low optical powers, where nonlinear effects are negligible, the propagation of the electric field remains entirely within the linear regime. The output field is a linear superposition of these guided modes. 
The resulting interference pattern, known as a speckle, has a superficially random and chaotic appearance.
However, this complexity is underpinned by a deterministic linear transformation, as the input wavefront and the output speckle pattern are connected by a fixed transmission matrix that characterizes the fiber. This deterministic nature is evidenced by the fact that the speckle pattern can be precisely predicted and even inverted using techniques such as wavefront shaping or artificial neural networks~\cite{borhani2018learning,rahmani2018multimode,moran2018deep}.

The statistical properties of speckle patterns in MMFs have been extensively characterized, primarily through the lens of probability density functions and spatial intensity correlations~\cite{goodman1975statistical,tsuji1984variation,imai1986statistical}. These studies have firmly established that under fully developed speckle conditions, the intensity statistics follow a negative exponential distribution, consistent with fully coherent random interference~\cite{goodman2007speckle}. 
Because this interference is highly phase-dependent, the resulting speckle pattern is extremely sensitive to external perturbations. This property underpins fiber specklegram sensors for physical variables such as strain and temperature, as well as high-resolution spectrometers~\cite{regez2009novel,fujiwara2018polymer,musin2016fiber,redding2013all,wang2023multimode,newaz2023machine}.
Furthermore, correlations in the speckle pattern have been leveraged to analyze mode coupling and propagation dynamics~\cite{popoff2010measuring,carpenter2015observation}. 
However, while these conventional statistical measures are well-established, they typically do not probe the multi-scale, higher-order statistical structure that is central to turbulence. To our knowledge, the application of ESS analysis to speckle patterns has not been previously reported.

In this work, we present the experimental observation of ESS in MMFs. We analyze the scaling of intensity structure functions from speckle patterns generated by four different laser wavelengths propagating through three MMFs with different core diameters. 
These experimental results are verified by numerical simulations. The observed scaling exponents, when compared against the third-order structure function, align with the universal scaling predicted by K41 theory for classical turbulence. 
This finding indicates that the universal scaling relations of ESS are not exclusive to nonlinear turbulent systems, but can also emerge from the complex interference inherent to linear wave transport.

\section{Experiment Setup}
\begin{figure}[htb]
    \centering
    \includegraphics{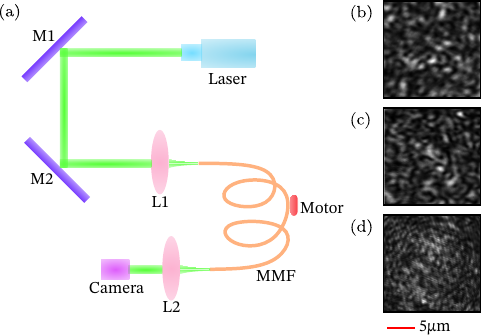}
    \caption{Experimental setup and representative speckle patterns. (a) Schematic of the optical setup used to generate and record speckle patterns. L: lens; M: mirror. (b-d) Representative speckle patterns measured at the output of 2-meter-long MMFs with core diameters of (b) $200\ {\rm{\mu m}}$, (c) $105\ {\rm{\mu m}}$, and (d) $50\ {\rm{\mu m}}$, respectively, using a $532\,\rm{nm}$ laser.}
    \label{fig:setup}
\end{figure}

The experimental setup for generating MMF speckle patterns is illustrated in Fig.~\ref{fig:setup}(a). A near-TEM\textsubscript{00} Gaussian beam from a semiconductor laser source was directed via two steering mirrors (M1, M2) and a collimating lens (L1) into the input facet of the step-index MMF (orange path). The output speckle pattern was collimated by a second lens (L2) and recorded using a CMOS camera.   To investigate the universality of ESS in speckle statistics, measurements were performed using four laser wavelengths ($\lambda = 532$, $660$, $780$, and $980~\text{nm}$) and three fiber core diameters ($D=50$, $105$, and $200\ {\rm{\mu m}}$).
All tested MMFs had a fixed length of $2\ \rm{m}$ to ensure sufficient mode mixing.

For each case, an ensemble for statistical analysis was created by collecting 100 independent speckle images. The variation between images was generated by weakly perturbing the fiber's local curvature, 
using a miniature vibrating motor ($10~\text{mm}$ diameter, $9000~\text{rpm}$) attached to the MMF. 
To enhance the sensitivity of the speckle variation to the mechanical vibration, a fiber with a thin external protective sleeve (total diameter $0.9~\text{mm}$) was used. 
The camera exposure time was set below $100\ \rm{\mu s}$ to ensure it was significantly shorter than the period of the mechanical vibration.

Representative speckle patterns for a $532\,\text{nm}$ source are shown in Fig.~\ref{fig:setup}(b-d) for core diameters of $200$, $105$, and $50\,{\rm{\mu m}}$, respectively. A clear morphological difference is observed between the patterns generated by the larger-core fibers ($200$ and $105\,{\rm{\mu m}}$) and the smaller-core fiber ($50\,{\rm{\mu m}}$).The speckles in (b) and (c) exhibit high contrast---defined as 
$C={\rm stdev}[\widetilde{I}]/{\rm mean}[\widetilde{I}]$---with measured 
values of $C = 0.47$ and $C = 0.51$, respectively, and are 
well-isolated, bearing a visual resemblance to the structures found in turbulent velocity fields. In contrast, the pattern for the $50\,{\rm{\mu m}}$ core in (d) shows  lower contrast $C=0.42$ and less distinct speckle grains.
This difference in speckle statistics is attributed to the reduced number of guided modes supported by the fiber with the smallest core diameter.
With fewer modes contributing to the interference, the resulting superposition is less complex.

\section{Extended Self-similarity}
\subsection{Structure Function}
The statistical analysis is based on the structure functions of the intensity field. For a two-dimensional speckle pattern $I(x,y)$, we define the intensity increment $\delta I_{ij} = I(x_i, y_i) - I(x_j, y_j)$ between two points separated by a distance  $l_{ij} = \sqrt{(x_i - x_j)^2 + (y_i - y_j)^2}$.

In analogy to Reynolds decomposition for turbulent velocity fields, the observed raw intensity $I_{\text{raw}}$ is decomposed into a mean component and fluctuations:
$I_{\text{raw}}(x,y) = \overline{I}(x,y) + \widetilde{I}(x,y)$.
Here, the mean intensity $\overline{I}$ is computed by averaging over the ensemble of 100 images within a given experimental dataset, isolating the fluctuating component $\widetilde{I}(x,y)$. This step is necessary as structure function analysis typically probes the statistics of fluctuations rather than the absolute intensity.

Since intensity is a scalar field, we employ the absolute structure function.
This function of order $p$ is calculated as the ensemble average of the $p$-th power of the absolute value of the fluctuation increments, conditional on a fixed separation $r$:
\begin{equation}
S_p(r) = \langle |\delta \widetilde{I}_{ij}|^p \rangle_{l_{ij} = r},
\end{equation}
where $\delta \widetilde{I}_{ij} = \widetilde{I}(x_i, y_i) - \widetilde{I}(x_j, y_j)$.

\begin{figure}[htb]
    \centering
    \includegraphics{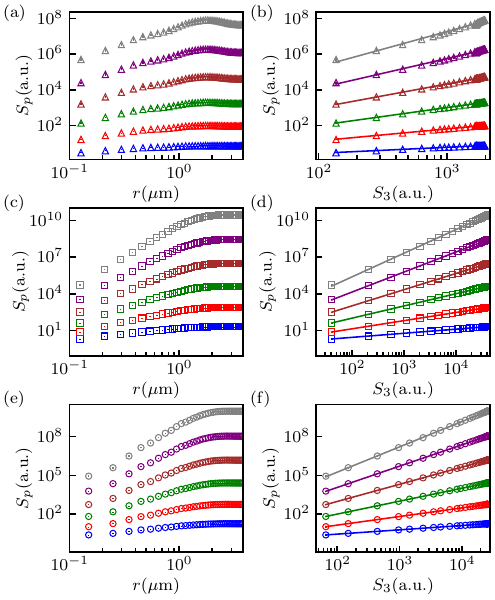}
    \caption{Structure functions and ESS for different fiber core diameters $D$ at $\lambda = 532$ nm. In each subplot, order $p$ increases from bottom to top. The top row (a, b) corresponds to $D = 50\,{\rm{\mu m}}$ (triangles), the middle row (c, d) to $D = 105\,{\rm{\mu m}}$ (squares), and the bottom row (e, f) to $D = 200\,{\rm{\mu m}}$ (circles). Left column (a, c, e): $S_p(r)$ versus spatial scale $r$. Right column (b, d, f): $S_p(r)$ versus $S_3(r)$. Solid lines are linear fits in log-log scale. Error bars represent $2\times$ standard errors of the mean.}
    \label{fig:S1_6}
\end{figure}



For the analysis, the intensity is treated as a dimensionless quantity, as our focus is on the scaling exponents, which are independent of physical units. The structure functions $S_p(r)$ for orders $p = 1, 2, ..., 6$ are plotted in Fig.~\ref{fig:S1_6}(a, c, e) for the $532\,\text{nm}$ source with MMF core diameters of $50$, $105$, and $200\,{\rm{\mu m}}$, respectively.

A common feature across all three cases is the saturation of $S_p(r)$ at separations $r > 2\,{\rm{\mu m}}$. For the larger core diameters ($105$ and $200\,{\rm{\mu m}}$), the structure functions suggest a possible power-law region within approximately $0.2\,{\rm{\mu m}} < r < 0.8\,{\rm{\mu m}}$. For the $50\,{\rm{\mu m}}$ core, two more limited scaling regions can be identified: $0.1\,{\rm{\mu m}} < r < 0.3\,{\rm{\mu m}}$ and $0.3\,{\rm{\mu m}} < r < 1\,{\rm{\mu m}}$.

The absence of a clean power law with large scale range in $S_p(r)$ is consistent with theoretical expectation. In a linear system dominated by mode interference, there is no inherent cascade process to establish a clear scale separation~\cite{aluie2011compressible} or a well-defined inertial range, in contrast to nonlinear turbulent systems.

This pattern—where conventional scaling with $r$ is absent, yet self-similarity across orders is pronounced—is consistently observed across all experimental cases. Detailed results for the other wavelengths are provided in Appendix~\ref{app:Sp_other}. The universal presence of this correlation among the $S_p(r)$ curves, despite the lack of a clean inertial range, strongly motivates the application of ESS to extract the underlying scaling relationships.

\subsection{the Extended Self-Similarity}



In classical turbulence, the inertial range over which conventional self-similarity holds is often limited at finite Reynolds numbers. The method of ESS circumvents this limitation by comparing structure functions of different orders against one another. This yields an extended scaling relationship of the form
\begin{equation}
    S_p(r) = \alpha_p \left[ S_3(r) \right]^{\beta_p},
\end{equation}
where $\alpha_p$ is a constant and $\beta_p$ is the ESS scaling exponent. In the absence of intermittency, $\beta_p$ follows the K41 prediction of $p/3$. 

\begin{figure}[tbh]
    \centering
    \includegraphics{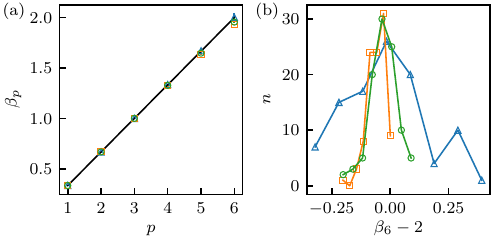}
    \caption{ ESS scalings.
    (a) Measured ESS scaling exponent $\beta_p$ as a function of order $p$ for the three fiber core diameters [$50\ \rm {\mu m}$ (triangle), $105\ \rm {\mu m}$ (square), and  $200\ \rm {\mu m}$ (circle)] with $532\,\rm{nm}$ laser. 
    The black line represents the classical K41 scaling relation $\beta_p = p/3$.
        (b) Histograms of the deviation $\beta_6$ from the K41 prediction ($\beta_6 = 2$), obtained from 100 independent speckle realizations for each core diameter.}
    \label{fig:ESS532}
\end{figure}

As shown in Fig.~\ref{fig:S1_6}(b,d,f), we apply this analysis to our speckle patterns by plotting $S_p$ versus $S_3$ on a log-log scale. In all cases, a clear linear relationship is observed across more than one decade, demonstrating the robust presence of ESS. The extracted exponents $\beta_p$, obtained from a linear fit, are summarized in Fig.~\ref{fig:ESS532}(a).

The results reveal that the low-order exponents ($p=1,2$) agree well with the classical K41 scaling $\beta_p = p/3$ (solid black line). However, increasing deviations from this linear relation emerge at higher orders ($p>4$).
Overall, the close alignment of the measured $\beta_p$ with the $p/3$ line indicates that the system exhibits very weak intermittency.

While the visual similarity between the larger-core speckle patterns [Fig.~\ref{fig:setup}(b-c)] and turbulent fields hints at the presence of ESS, its empirical confirmation in a linear, deterministic system is a useful conceptual extension. This result demonstrates that the statistical signatures captured by ESS are not exclusive to nonlinear chaotic systems but can also emerge from complex linear wave interference.

The emergence of ESS can be interpreted as a consequence of a mode-mixing process that shares a statistical similarity with the direct energy cascade in nonlinear wave turbulence. Here, linear propagation in a bent or disordered waveguide effectively redistributes energy (or modal amplitude) across transverse wavenumbers, generating complex interference that mimics the multi-scale coupling of a cascade. This linear mode coupling mechanism is explored numerically in Sec.~\ref{sec:numerical}.

The fundamental reason why this linear process yields scaling exponents so closely aligned with the K41 scalings remains an open theoretical question. However, the deterministic and well-characterized nature of linear wave transport makes this system a promising and potentially simpler testbed for probing the essential conditions for ESS, potentially opening a new path to a deeper understanding of its origins.



To determine whether ESS scaling is a consistent property of individual speckle realizations, we computed the ESS exponents $\beta_p$ for each image in the ensemble. The deviation from the K41 prediction for the sixth-order exponent, $\beta_6 - 2$, serves as a metric to quantify the degree of intermittency and the agreement with theory.

The distributions of $\beta_6 - 2$ are shown in Fig.~\ref{fig:ESS532}(b). All distributions are centered near zero, indicating consistent agreement with K41 scaling across the ensemble. However, distinct trends emerge based on core diameter. For the larger-core fibers ($105\,{\rm{\mu m}}$ and $200\,{\rm{\mu m}}$), the distributions are narrow but systematically shifted to negative values, indicating a small, consistent bias toward higher intermittency. In contrast, the distribution for the $50\,{\rm{\mu m}}$ core fiber is broader [as also reflected in the larger error bars in Fig.~\ref{fig:ESS532}(a)] but is centered closer to zero, suggesting a lower average level of intermittency despite greater variability between individual speckle patterns.


In summary, we demonstrate the existence of ESS in speckle patterns generated by linear propagation through MMFs, with scaling exponents that follow the universal K41 scalings (see Appendix.~\ref{app:Sp_other} for all tested cases). 


\subsection{Intermittency of $S_p(r)$ under ESS}

The more complete Kolmogorov–Obukhov 1962 (KO62) theory~\cite{kolmogorov1962refinement,obukhov1962some} refines the K41 framework by introducing an intermittency correction to account for the non-Gaussian fluctuations in the energy dissipation rate. This correction becomes significant at higher orders $p$ and is often modeled under the assumption of a log-normal distribution for the dissipation rate. Within this framework, the deviation of the measured ESS scaling exponents from the K41 prediction can be parameterized by a quadratic form
\begin{equation}
\beta_p - \frac{p}{3} = \kappa \, p(p - 3),
\label{eq:kappa}
\end{equation}
where the parameter $\kappa$ quantifies the strength of intermittency in the system. 

\begin{figure}[tbh]
    \centering
    \includegraphics{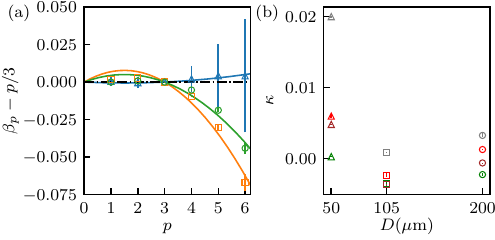}
    \caption{Intermittency analysis of fiber speckle patterns.
    (a) Deviation of the ESS scaling exponents from the K41 prediction, $\beta_p - p/3$, plotted against order $p$ for $\lambda = 532\,\text{nm}$. Data for core diameters of $200\,{\rm{\mu m}}$, $105\,{\rm{\mu m}}$, and $50\,{\rm{\mu m}}$ are shown as circles, squares, and triangles, respectively. The solid lines are parabolic fits of the form $\kappa \, p(p-3)$, from which the intermittency strength $\kappa$ is extracted. (b) Summary of the fitted $\kappa$ values for all experimental conditions (4 wavelengths $\times$ 3 core diameters). Colors denote wavelength: $532\,\text{nm}$ (green), $660\,\text{nm}$ (red), $780\,\text{nm}$ (brown), and $980\,\text{nm}$ (gray). Error bars represent the standard error from the fitting procedure.
 }
    \label{fig:Intermittency}
\end{figure}

While the KO62 theory originates in the context of hydrodynamic turbulence, the parabolic form serves as a robust empirical fit to characterize the intermittency in our MMF speckles.
Figure~\ref{fig:Intermittency}(a) shows this deviation $\beta_p - p/3$ for the $532\,\text{nm}$ cases. We observe that for the larger core diameters ($105$ and $200\,{\rm{\mu m}}$), the derived $\kappa < 0$, a signature also typical in classical fluid turbulence. In contrast, the small core diameter case ($50\,{\rm{\mu m}}$) yields a positive $\kappa$, which implies a different statistical structure for the fluctuations.

We fitted the intermittency exponent $\kappa$ via Eq.~(\ref{eq:kappa}) for all experimental conditions (see Appendix~\ref{app:Sp_other}) and summarize the results in Fig.~\ref{fig:Intermittency}(b). The $\kappa$ values for the $D=50\,{\rm{\mu m}}$ fiber are constantly positive and notably larger in magnitude than those for fibers with larger cores. Among all configurations tested, the $D=105\,{\rm{\mu m}}$ fiber exhibits the smallest $\kappa$ values, closest to the zero baseline of K41 scaling. For a given core diameter $D$, the measured $\kappa$ varies with wavelength $\lambda$, generally descending in the order $\kappa_{980\,\text{nm}} > \kappa_{660\,\text{nm}} > \kappa_{780\,\text{nm}} > \kappa_{532\,\text{nm}}$. The strongest intermittency effect is observed for the $980\,\text{nm}$ source in the $D=50\,{\rm{\mu m}}$ fiber, the configuration that supports the fewest guided modes.

Overall, across the wide range of core diameters and wavelengths investigated, the magnitude of $\kappa$ remains small ($|\kappa| \ll 0.02$), indicating that the overall intermittency in these linear speckle systems is weak.

\section{Numerical validation}\label{sec:numerical}

\subsection{Speckle generation using coupled mode theory}

We numerically model a step-index optical fiber with a diameter of $105\,{\rm{\mu m}}$, a core refractive index of $n_1=1.47$, and a cladding index of $n_2=1.46$. The operating wavelength is $\lambda=980\,\rm{nm}$. Under the weakly guiding approximation~\cite{gloge1971weakly}, the guided modes in cylindrical coordinates take the form $\psi(r,\phi)e^{i(\omega t-\beta z)}$, where $\omega=2\pi c/\lambda$ and the propagation constant $\beta$ lies in the interval $[2\pi n_2/\lambda,2\pi n_1/\lambda]$.
By solving for the linearly polarized (LP) modes~\cite{ghatak1998introduction} $\Psi_{lm}$—labeled by the angular index $l$ and the radial index $m$—we identify a total of $N=802$ guided modes. In the following we use a single index $q=\{l,m\}$ to label different modes. A detailed discussion of the mode calculation is provided in the appendix~\ref{app:LP}.

We employ coupled mode theory (CMT) to model optical fiber propagation and generate speckle patterns at the fiber output~\cite{snyder1972coupled}. The electric field within the fiber can be expressed as a linear superposition of LP modes:
\begin{equation}
E(z, r, \phi) = \sum_{q} a_{q}(z) \Psi_{q}(r, \phi),
\end{equation}
where $\Psi_q(r, \phi)$ denotes the transverse field distribution of the $q$-th mode, and $a_q(z)$ represents its complex amplitude.
The evolution of mode amplitudes along the propagation direction $z$ is governed by the equation:
\begin{equation}
\frac{d a_q(z)}{dz} + i \beta_q a_q(z) = i \sum_{s}^{N} C_{qs}(z) a_s(z).
\label{eq:cmt}
\end{equation}
Here, $\beta_q$ is the propagation constant of mode $q$, and $C_{qs}(z)$ describes the coupling between modes $q$ and $s$ due to fiber imperfections. Under the normalization condition
\begin{equation}
\int \frac{\beta_q}{\omega\mu_0} |\Psi_q|^2 \, r \, dr \, d\phi = 1,
\end{equation}
the coupling coefficient is given by
\begin{equation}
C_{qs}(z) = \frac{\omega}{2} \int \Delta\epsilon(r, \phi, z) \Psi_q \Psi_s^* \, r \, dr \, d\phi,
\end{equation}
where $\Delta\epsilon(r, \phi, z)$ represents the deviation of the dielectric constant from that of a perfect, straight cylindrical fiber.

We consider the weak-coupling regime, characterized by $\max_{q \neq s} |C_{qs}| \ll \min_q |\beta_q|$,
ensuring that mode coupling occurs on a length scale much longer than the wavelength. For simplicity, we assume $z$-independent coupling coefficients $C_{qs}$. The off-diagonal elements of the coupling matrix are modeled as independent random complex numbers, with real and imaginary parts uniformly distributed between $-\min_q |\beta_q|/10000$ and $\min_q |\beta_q|/10000$. The matrix is constrained to be Hermitian, i.e., $C_{pq} = C_{qp}^*$, which ensures power conservation.
The numerical integration of Eq.~\eqref{eq:cmt} is performed using a fifth-order explicit Runge–Kutta method with fourth-order error control~\cite{dormand1980family,shampine1986some} and an adaptive step size.

\begin{figure}[htb]
    \centering
    \includegraphics{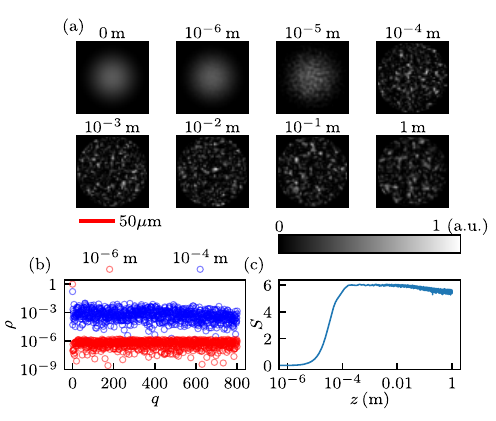}
    \caption{Numerical simulation of optical mode evolution in an MMF. (a) Simulated intensity patterns $I(z,x,y)=|E(z,x,y)|^{2}$ at propagation distances $z = 0, 10^{-6}, 10^{-5}, 10^{-4}, 10^{-3}, 10^{-2}, 10^{-1}, 1~\mathrm{m}$. The simulation parameters are $\lambda = 980\,\mathrm{nm}$ and  $D=105\,\mathrm{\mu m}$. (b) Distribution of mode weights $\rho_{q}$ at $10^{-6}\,\mathrm{m}$ (red), and $10^{-4}\,\mathrm{m}$ (blue). (c) Evolution of the von Neumann entropy $S(z)$ as a function of propagation distance.}
    \label{fig:numerical_evolve}
\end{figure}

Figure~\ref{fig:numerical_evolve}(a) displays the evolution of the intensity pattern $I(z,x,y) = |E(z,x,y)|^{2}$ at several propagation distances $z$. The initial pattern $I(0,x,y)$ corresponds to a pure $\mathrm{LP}_{01}$ mode. As propagation proceeds, modal diffusion becomes apparent: by $z = 10^{-5}\,\mathrm{m}$, speckles emerge while the intensity remains dominated by radial modes with $m=1$. At $z = 10^{-4}\,\mathrm{m}$, the intensity fills the entire fiber core with a fully developed speckle pattern. This corresponds to a near-equilibrium mixture of mode weights $\rho_{q}=|a_q|^2$, illustrated by the blue dots in Fig.~\ref{fig:numerical_evolve}(b), in contrast to the early-stage ($z=10^{-6}\,\mathrm{m}$) distribution dominated by only a few modes.

To quantify the degree of mode mixing, we compute the von Neumann entropy $S = -\sum_{q} \rho_{q} \ln(\rho_{q})$~\cite{bengtsson2017geometry}. The resulting entropy $S(z)$, shown in Fig.~\ref{fig:numerical_evolve}(c), saturates at approximately $z = 10^{-4}\,\mathrm{m}$, consistent with the visual transition observed in Fig.~\ref{fig:numerical_evolve}(a).

\subsection{ESS for intensity pattern}

We examine the ESS for numerically generated speckle patterns. Rather than assembling experimental ensembles from fibers of fixed length, we construct a dataset from a single propagation sequence. This dataset comprises 100 patterns, each separated by $\Delta z = 10^{-5}\,\mathrm{m}$ in the region around $z = 10^{-2}\,\mathrm{m}$, where the mode mixture has reached a fully developed state, as evidenced by the saturated entropy in Fig.~\ref{fig:numerical_evolve}(c).

\begin{figure}[htb]
    \centering
    \includegraphics{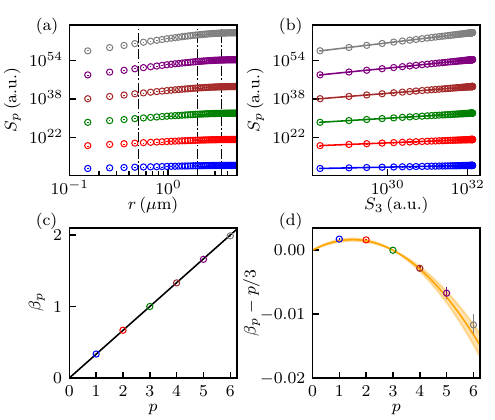}
    \caption{ESS analysis of numerically generated speckle patterns from linear mode mixing. (a) Structure functions $S_p(r)$ for $p=1,2,3...,6$, color-coded as shown on the
x-axis in panel (c), saturate at $r \approx 3.5\,\mathrm{\mu m}$. Vertical dashed lines mark three scaling regions. (b) ESS plot of $S_p(r)$ versus $S_3(r)$ in log-log scale. (c) Measured scaling exponents $\beta_p$ (points) exhibit near-perfect agreement with the K41 prediction $\beta_p = p/3$ (solid black line). (d) Deviation from K41 scaling, $\beta_p - p/3$, fitted by the KO62 intermittency model (orange curve; shaded region indicates $2\sigma$ uncertainty). }
    \label{fig:ESS_numerical}
\end{figure}

Using the same analysis method as before, we first obtain the mean intensity pattern. We then calculate the fluctuations of individual speckle patterns, from which we derive the structure functions $S_p(r)$. The result for $S_p(r)$ is shown in Fig.~\ref{fig:ESS_numerical}(a); it saturates at approximately $r \approx 3.5\,\mathrm{\mu m}$, consistent with the experimental observation.

We identify three distinct scaling regions, marked by the vertical black dashed lines in Fig.~\ref{fig:ESS_numerical}(a): $r < 0.5\,\mathrm{\mu m}$, $0.5\,\mathrm{\mu m} < r < 2.0\,\mathrm{\mu m}$, and $2.0\,\mathrm{\mu m} < r < 3.5\,\mathrm{\mu m}$. Because the latter two nontrivial scaling ranges are relatively narrow, we apply the ESS analysis, plotting $S_p(r)$ versus $S_3(r)$ in Fig.~\ref{fig:ESS_numerical}(b). The resulting scaling exponents $\beta_p$, obtained from linear fits in the ESS log-log representation, are shown in Fig.~\ref{fig:ESS_numerical}(c). They are in excellent agreement with the K41 prediction $\beta_p = p/3$, indicated by the solid black line.

Figure~\ref{fig:ESS_numerical}(d) plots the deviation $\beta_p - p/3$ and compares it to a fit based on the KO62 intermittency model [Eq.~\eqref{eq:kappa}] (orange curve). The surrounding shaded region represents the $2\sigma$ uncertainty of the fit. The fitted intermittency strength is $\kappa=-7.3(4) \times 10^{-4}$, indicating very weak intermittency in the system. We therefore confirm that ESS scaling can emerge in a purely linear mode-mixing process.

\section{Conclusion}
In this work, we demonstrate the emergence of ESS in a purely linear optical system: the speckle patterns generated by coherent light propagation in MMFs. Through systematic experiments across multiple wavelengths ($532$, $660$, $780$, and $980$\,nm) and fiber core diameters ($50$, $105$, and $200\,{\rm{\mu m}}$), we show that the intensity structure functions of these speckles exhibit a robust extended scaling range when analyzed via the ESS procedure. The extracted scaling exponents align with the universal $\beta_p = p/3$ relation predicted by K41 theory for classical turbulence. This finding is further corroborated by numerical simulations, confirming that the observed scaling is an intrinsic property of the linear interference process.

Our results establish two key points. First, they show that the universal statistical signatures captured by ESS are not exclusive to systems governed by nonlinear dynamics or energy cascades. The complex, multi-scale interference of a large but finite number of modes in a linear waveguide is sufficient to produce the same scaling relations.
Second, using the intermittency analysis framework of the KO62 theory, we quantified weak but systematic deviations from K41 scaling. The sign and magnitude of the intermittency strength $\kappa$ were found to depend systematically on the fiber core diameter and laser wavelength, being most pronounced for the fiber supporting the fewest guided modes.

The ESS framework is applicable to linear, deterministic systems, thereby revealing a deeper universality in the statistics of complex wavefields. It positions linear optical speckle as a new and well-controlled testbed for studying universal scaling phenomena. More broadly, our findings suggest that other linear disordered systems exhibiting complex interference—such as in acoustics or quantum matter-waves—may also host analogous universal scaling, opening new avenues for interdisciplinary research into complex system behavior.

This study employed step-index fibers. A natural and important extension is to investigate ESS in speckle patterns from graded-index (GRIN) fibers.
Comparing the resulting scaling exponents and intermittency corrections will test the robustness of universal scaling across different waveguide geometries.

In the nonlinear regime, GRIN fibers exhibit phenomena such as Kerr-induced spatial self-cleaning~\cite{krupa2017spatial}. This process, where the beam condenses into a fundamental-like mode, can be interpreted as an inverse cascade in a 2D nonlinear turbulent fluid system~\cite{PhysRevLett.122.103902}.
Applying the ESS and intermittency analysis developed here to the transition from disordered speckle to the ordered state could provide powerful statistical insights into the underlying nonlinear physics and turbulent dynamics.

Our results demonstrate that the intermittency strength $\kappa$ varies sensitively and systematically with the input wavelength. This establishes a foundational principle for a new class of statistical spectrometers or wavemeters. By calibrating the higher-order statistical fingerprints (e.g., $\kappa$ or $\alpha_p$) of a speckle pattern against wavelength, one could leverage a simple MMF and a camera for precise optical sensing.
\begin{acknowledgments}
The authors thank Gaoqing Meng for carefully reading the manuscript.
This work was partially supported by the Research and Practice Project of Higher Education Teaching Reform of Hebei Province (No.2025GJJG412).
\end{acknowledgments}

\begin{appendix}
\numberwithin{equation}{section}

\section{Structure functions and intermittency for various laser wavelengths}\label{app:Sp_other}

This appendix presents the supplementary scaling and intermittency analyses for the laser wavelengths of $660\,\text{nm}$, $780\,\text{nm}$, and $980\,\text{nm}$. For each wavelength, two figures are provided: the first shows the structure functions $S_p(r)$ versus scale $r$ alongside their ESS representation $S_p$ versus $S_3$; the second presents the resulting ESS scaling exponents $\beta_p$ and the corresponding intermittency fit $\kappa p(p-3)$. 

\begin{figure}[htb]
    \centering
    \includegraphics{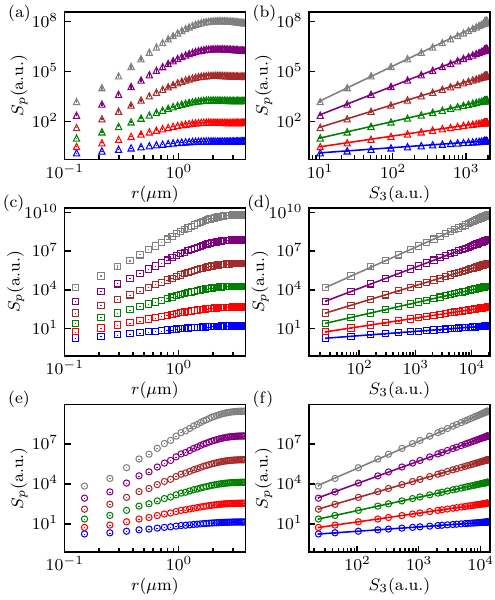}
    \caption{Structure functions and ESS for different fiber core diameters $D$ at $\lambda = 660$ nm. In each subplot, order $p$ increases from bottom to top. The top row (a, b) corresponds to $D = 50\,{\rm{\mu m}}$ (triangles), the middle row (c, d) to $D = 105\,{\rm{\mu m}}$ (squares), and the bottom row (e, f) to $D = 200\,{\rm{\mu m}}$ (circles). Left column (a, c, e): $S_p(r)$ versus spatial scale $r$. Right column (b, d, f): $S_p(r)$ versus $S_3(r)$. Solid lines are linear fits in the log-log scale. Error bars represent $2\times$ standard errors of the mean.}
    \label{fig:S1_6_660}
\end{figure}

\begin{figure}[htb]
    \centering
    \includegraphics{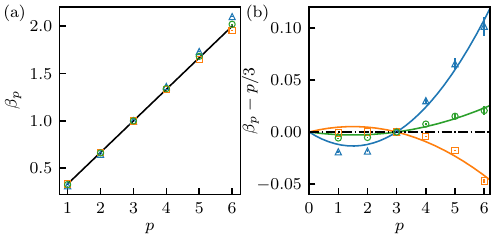}
    \caption{ESS scalings with the $660\,\rm{nm}$ laser.
    Data for core diameters of $200\,{\rm{\mu m}}$, $105\,{\rm{\mu m}}$, and $50\,{\rm{\mu m}}$ are shown as circles, squares, and triangles, respectively.
    (a) $\beta_p$ as a function of order $p$ for the three fiber core diameters. The black line represents the K41 scalings $\beta_p = p/3$.
        (b) Deviation of the $\beta_p$ from the K41 prediction, $\beta_p - p/3$, plotted against order $p$.  The solid lines are fits of the form $\kappa \, p(p-3)$.}
    \label{fig:ESS660}
\end{figure}

\begin{figure}[htb]
    \centering
    \includegraphics{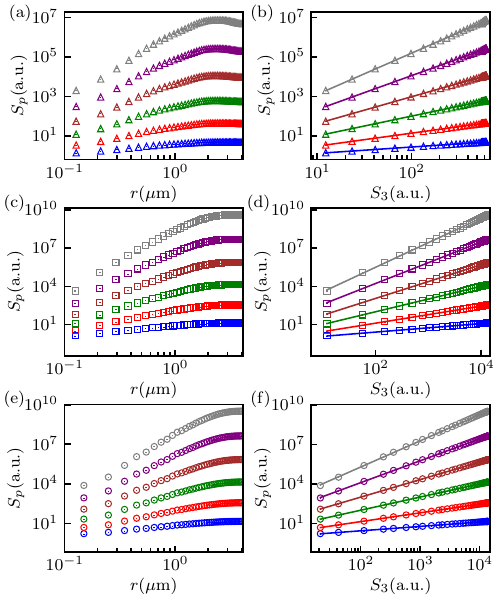}
    \caption{Structure functions and ESS for different fiber core diameters $D$ at $\lambda = 780$ nm. In each subplot, order $p$ increases from bottom to top. The top row (a, b) corresponds to $D = 50\,{\rm{\mu m}}$ (triangles), the middle row (c, d) to $D = 105\,{\rm{\mu m}}$ (squares), and the bottom row (e, f) to $D = 200\,{\rm{\mu m}}$ (circles). Left column (a, c, e): $S_p(r)$ versus spatial scale $r$. Right column (b, d, f): $S_p(r)$ versus $S_3(r)$. Solid lines are linear fits in the log-log scale. Error bars represent $2\times$ standard errors of the mean.}
    \label{fig:S1_6_780}
\end{figure}

\begin{figure}[tbh]
    \centering
    \includegraphics{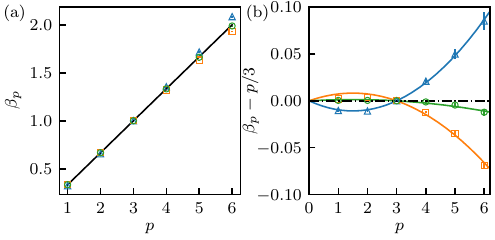}
    \caption{ ESS scalings with the $780\,\rm{nm}$ laser.
    Data for core diameters of $200\,{\rm{\mu m}}$, $105\,{\rm{\mu m}}$, and $50\,{\rm{\mu m}}$ are shown as circles, squares, and triangles, respectively.
    (a) $\beta_p$ as a function of order $p$ for the three fiber core diameters. The black line represents the K41 scalings $\beta_p = p/3$.
        (b) Deviation of the $\beta_p$ from the K41 prediction, $\beta_p - p/3$, plotted against order $p$.  The solid lines are fits of the form $\kappa \, p(p-3)$.}
    \label{fig:ESS780}
\end{figure}

\begin{figure}[htb]
    \centering
    \includegraphics{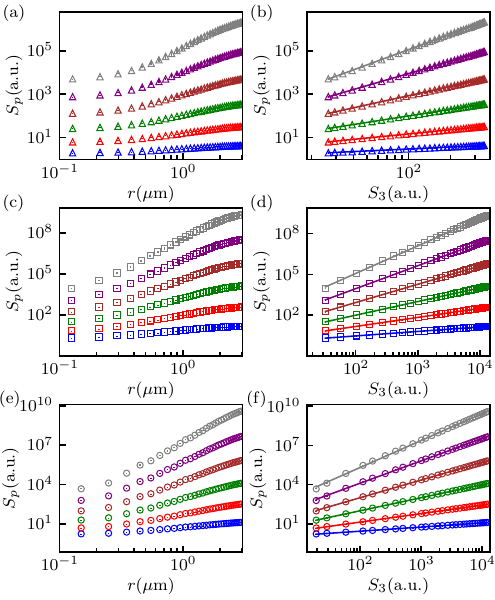}
    \caption{Structure functions and ESS for different fiber core diameters $D$ at $\lambda = 980$ nm. In each subplot, order $p$ increases from bottom to top. The top row (a, b) corresponds to $D = 50\,{\rm{\mu m}}$ (triangles), the middle row (c, d) to $D = 105\,{\rm{\mu m}}$ (squares), and the bottom row (e, f) to $D = 200\,{\rm{\mu m}}$ (circles). Left column (a, c, e): $S_p(r)$ versus spatial scale $r$. Right column (b, d, f): $S_p(r)$ versus $S_3(r)$. Solid lines are linear fits to the ESS scaling regimes. Error bars represent $2\times$ standard errors of the mean.}
    \label{fig:S1_6_980}
\end{figure}

\begin{figure}[htb]
    \centering
    \includegraphics{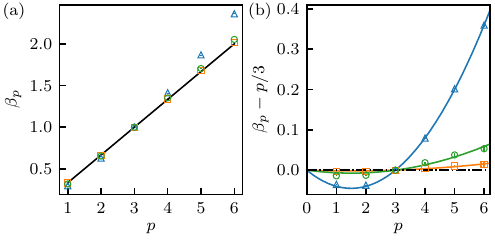}
    \caption{ESS scalings with the $980\,\rm{nm}$ laser.
    Data for core diameters of $200\,{\rm{\mu m}}$, $105\,{\rm{\mu m}}$, and $50\,{\rm{\mu m}}$ are shown as circles, squares, and triangles, respectively.
    (a) $\beta_p$ as a function of order $p$ for the three fiber core diameters. The black line represents the K41 scalings $\beta_p = p/3$.
        (b) Deviation of the $\beta_p$ from the K41 prediction, $\beta_p - p/3$, plotted against order $p$.  The solid lines are fits of the form $\kappa \, p(p-3)$.}
    \label{fig:ESS980}
\end{figure}

Specifically, for $660\,\text{nm}$, the structure functions and ESS representations are shown in Fig.~\ref{fig:S1_6_660}, while the scaling exponents and intermittency fit are shown in Fig.~\ref{fig:ESS660}. Similarly, the corresponding figures for $780\,\text{nm}$ are Fig.~\ref{fig:S1_6_780} and Fig.~\ref{fig:ESS780}, and for $980\,\text{nm}$ are Fig.~\ref{fig:S1_6_980} and Fig.~\ref{fig:ESS980}. Together with the $532\,\text{nm}$ data discussed in the main text, this full set of measurements forms the basis for the wavelength and diameter-dependent trends in intermittency parameters summarized in Fig.~\ref{fig:Intermittency}(b).


\section{Modal analysis in optical fibers}\label{app:LP}

The multimode fibers used in this study are step-index fibers with a radially symmetric refractive index profile:
\begin{equation}
    n(r) = \begin{cases}
        n_1, & 0 < r < a, \quad \text{(core)} \\
        n_2, & r > a, \quad \text{(cladding)}
    \end{cases}
\end{equation}
where $a$ is the core radius. For standard fibers, the core and cladding indices are very close (e.g., $n_1 = 1.47$, $n_2 = 1.46$), validating the weakly guiding approximation. This approximation implies that the guided modes are nearly transverse electromagnetic waves, allowing the problem to be reduced to solving a scalar Helmholtz equation for a transverse field component $\Psi$ (representing $E_x$ or $E_y$):
\begin{equation}
    \nabla^2\Psi = \epsilon_0\mu_0n^2(r)\partial_t^2\Psi .
    \label{eq:scalar_Helmholtz}
\end{equation}

Exploiting cylindrical symmetry, we seek solutions propagating along the fiber axis ($z$-direction) of the form
\begin{equation}
    \Psi(r,\phi,z,t) = \psi(r,\phi)e^{i(\omega t - \beta z)},
\end{equation}
where $\beta$ is the propagation constant. Substituting this ansatz into Eq.~\eqref{eq:scalar_Helmholtz} yields
\begin{equation}
    \partial_r^2\psi + \frac{1}{r}\partial_r\psi + \frac{1}{r^2}\partial_{\phi}^2\psi + \left[ k_0^2 n^2(r) - \beta^2 \right]\psi = 0,
    \label{eq:reduced_helmholtz}
\end{equation}
with $k_0^2 = \omega^2\epsilon_0\mu_0$.

Applying separation of variables, $\psi(r,\phi) = R(r)\Phi(\phi)$, and imposing the angular periodicity condition $\Phi(\phi+2\pi) = \Phi(\phi)$, leads to
\begin{equation}
    \Phi(\phi) = e^{il\phi}, \quad l = 0, \pm 1, \pm 2, \dots .
\end{equation}
The radial function $R(r)$ then satisfies
\begin{equation}
    r^2R'' + rR' + \left\{ \left[ k_0^2 n^2(r) - \beta^2 \right] r^2 - l^2 \right\}R = 0.
    \label{eq:radial_eq}
\end{equation}

For guided modes, the propagation constant $\beta$ must lie in the range $k_0 n_2 < \beta < k_0 n_1$. This ensures an oscillatory solution in the core ($k_0^2 n_1^2 - \beta^2 > 0$) and an evanescent, decaying solution in the cladding ($k_0^2 n_2^2 - \beta^2 < 0$). The physical solutions, finite at $r=0$ and as $r \to \infty$, are
\begin{equation}
    R(r) = \begin{cases}
        A\, J_l\left( U r/a \right), & r < a \\
        A\, \dfrac{K_l\left( W r/a \right)}{K_l(W)}, & r > a
    \end{cases}
    \label{eq:radial_solution}
\end{equation}
where $A$ is a normalization constant, $J_l$ is the Bessel function of the first kind, $K_l$ is the modified Bessel function of the second kind, and the dimensionless parameters are defined as
\begin{equation}
    U = a\sqrt{k_0^2 n_1^2 - \beta^2}, \qquad W = a\sqrt{\beta^2 - k_0^2 n_2^2}.
\end{equation}

The continuity of both the field and its radial derivative at the core-cladding interface ($r=a$) leads to the characteristic equation that determines the allowed eigenvalues $\beta$ for a given azimuthal order $l$:
\begin{equation}
    U \frac{J_{l+1}(U)}{J_l(U)} = W \frac{K_{l+1}(W)}{K_l(W)}.
    \label{eq:characteristic_eq}
\end{equation}
This equation must be solved numerically for $\beta$ given $l$ and the fiber parameters ($a$, $n_1$, $n_2$, $k_0$). Since $K_l(W) > 0$ and $K_{l+1}(W)/K_l(W)$ is positive and monotonically decreasing for $W>0$, solutions for a given $l$ correspond approximately to the roots of $J_l(U) \approx 0$. This provides an efficient initial guess for the numerical root-finding procedure. Each solution $(l, \beta)$ corresponds to a guided LP$_{lm}$ mode, where the integer $m$ indexes the radial solutions in order of increasing $\beta$.

In section.~\ref{sec:numerical} we consider the wavelength $\lambda=980\,\rm{nm}$ with the fiber core radius $a=52.5\,{\rm{\mu m}}$.
$\beta$ should be close to the zeros of $J_l$, i.e., $J_l(\sqrt{k_0^2 n_1^2-\beta^2}\ a)\approx0$.
Using the guiding mode condition, $\sqrt{k_0^2 n_1^2-\beta^2}\ a\in\left[0,k_0a\sqrt{n_1^2-n_2^2}\right]=[0,57.616]$.
This determines the maximal $l$ in the system, since at least one zero should be within this interval for the $J_l$.
The first zero of $J_{50}$ and $J_{51}$ are $57.117$ and $58.160$, respectively. Hence, with a maximum $l=50$, we find the total number of guiding modes to be $N = 802$.
\end{appendix}

\input{main.bbl}

\end{document}

%% file: main.bbl
%

%% file: main.bbl
\begin{thebibliography}{51}%
\makeatletter
\providecommand \@ifxundefined [1]{%
 \@ifx{#1\undefined}
}%
\providecommand \@ifnum [1]{%
 \ifnum #1\expandafter \@firstoftwo
 \else \expandafter \@secondoftwo
 \fi
}%
\providecommand \@ifx [1]{%
 \ifx #1\expandafter \@firstoftwo
 \else \expandafter \@secondoftwo
 \fi
}%
\providecommand \natexlab [1]{#1}%
\providecommand \enquote  [1]{``#1''}%
\providecommand \bibnamefont  [1]{#1}%
\providecommand \bibfnamefont [1]{#1}%
\providecommand \citenamefont [1]{#1}%
\providecommand \href@noop [0]{\@secondoftwo}%
\providecommand \href [0]{\begingroup \@sanitize@url \@href}%
\providecommand \@href[1]{\@@startlink{#1}\@@href}%
\providecommand \@@href[1]{\endgroup#1\@@endlink}%
\providecommand \@sanitize@url [0]{\catcode `\\12\catcode `\$12\catcode `\&12\catcode `\#12\catcode `\^12\catcode `\_12\catcode `\%12\relax}%
\providecommand \@@startlink[1]{}%
\providecommand \@@endlink[0]{}%
\providecommand \url  [0]{\begingroup\@sanitize@url \@url }%
\providecommand \@url [1]{\endgroup\@href {#1}{\urlprefix }}%
\providecommand \urlprefix  [0]{URL }%
\providecommand \Eprint [0]{\href }%
\providecommand \doibase [0]{https://doi.org/}%
\providecommand \selectlanguage [0]{\@gobble}%
\providecommand \bibinfo  [0]{\@secondoftwo}%
\providecommand \bibfield  [0]{\@secondoftwo}%
\providecommand \translation [1]{[#1]}%
\providecommand \BibitemOpen [0]{}%
\providecommand \bibitemStop [0]{}%
\providecommand \bibitemNoStop [0]{.\EOS\space}%
\providecommand \EOS [0]{\spacefactor3000\relax}%
\providecommand \BibitemShut  [1]{\csname bibitem#1\endcsname}%
\let\auto@bib@innerbib\@empty
\bibitem [{\citenamefont {Meneveau}\ and\ \citenamefont {Katz}(2000)}]{meneveau2000scale}%
  \BibitemOpen
  \bibfield  {author} {\bibinfo {author} {\bibfnamefont {C.}~\bibnamefont {Meneveau}}\ and\ \bibinfo {author} {\bibfnamefont {J.}~\bibnamefont {Katz}},\ }\href@noop {} {\bibfield  {journal} {\bibinfo  {journal} {Annual Review of Fluid Mechanics}\ }\textbf {\bibinfo {volume} {32}},\ \bibinfo {pages} {1} (\bibinfo {year} {2000})}\BibitemShut {NoStop}%
\bibitem [{\citenamefont {Alexakis}\ and\ \citenamefont {Biferale}(2018)}]{alexakis2018cascades}%
  \BibitemOpen
  \bibfield  {author} {\bibinfo {author} {\bibfnamefont {A.}~\bibnamefont {Alexakis}}\ and\ \bibinfo {author} {\bibfnamefont {L.}~\bibnamefont {Biferale}},\ }\href@noop {} {\bibfield  {journal} {\bibinfo  {journal} {Physics Reports}\ }\textbf {\bibinfo {volume} {767}},\ \bibinfo {pages} {1} (\bibinfo {year} {2018})}\BibitemShut {NoStop}%
\bibitem [{\citenamefont {Marino}\ and\ \citenamefont {Sorriso-Valvo}(2023)}]{marino2023scaling}%
  \BibitemOpen
  \bibfield  {author} {\bibinfo {author} {\bibfnamefont {R.}~\bibnamefont {Marino}}\ and\ \bibinfo {author} {\bibfnamefont {L.}~\bibnamefont {Sorriso-Valvo}},\ }\href@noop {} {\bibfield  {journal} {\bibinfo  {journal} {Physics Reports}\ }\textbf {\bibinfo {volume} {1006}},\ \bibinfo {pages} {1} (\bibinfo {year} {2023})}\BibitemShut {NoStop}%
\bibitem [{\citenamefont {Jiang}\ \emph {et~al.}(2019)\citenamefont {Jiang}, \citenamefont {Xie}, \citenamefont {Zhou},\ and\ \citenamefont {Sornette}}]{jiang2019multifractal}%
  \BibitemOpen
  \bibfield  {author} {\bibinfo {author} {\bibfnamefont {Z.-Q.}\ \bibnamefont {Jiang}}, \bibinfo {author} {\bibfnamefont {W.-J.}\ \bibnamefont {Xie}}, \bibinfo {author} {\bibfnamefont {W.-X.}\ \bibnamefont {Zhou}},\ and\ \bibinfo {author} {\bibfnamefont {D.}~\bibnamefont {Sornette}},\ }\href@noop {} {\bibfield  {journal} {\bibinfo  {journal} {Reports on Progress in Physics}\ }\textbf {\bibinfo {volume} {82}},\ \bibinfo {pages} {125901} (\bibinfo {year} {2019})}\BibitemShut {NoStop}%
\bibitem [{\citenamefont {Benzi}\ \emph {et~al.}(1993{\natexlab{a}})\citenamefont {Benzi}, \citenamefont {Ciliberto}, \citenamefont {Tripiccione}, \citenamefont {Baudet}, \citenamefont {Massaioli},\ and\ \citenamefont {Succi}}]{benzi1993extended}%
  \BibitemOpen
  \bibfield  {author} {\bibinfo {author} {\bibfnamefont {R.}~\bibnamefont {Benzi}}, \bibinfo {author} {\bibfnamefont {S.}~\bibnamefont {Ciliberto}}, \bibinfo {author} {\bibfnamefont {R.}~\bibnamefont {Tripiccione}}, \bibinfo {author} {\bibfnamefont {C.}~\bibnamefont {Baudet}}, \bibinfo {author} {\bibfnamefont {F.}~\bibnamefont {Massaioli}},\ and\ \bibinfo {author} {\bibfnamefont {S.}~\bibnamefont {Succi}},\ }\href@noop {} {\bibfield  {journal} {\bibinfo  {journal} {Physical review E}\ }\textbf {\bibinfo {volume} {48}},\ \bibinfo {pages} {R29} (\bibinfo {year} {1993}{\natexlab{a}})}\BibitemShut {NoStop}%
\bibitem [{\citenamefont {Benzi}\ \emph {et~al.}(1993{\natexlab{b}})\citenamefont {Benzi}, \citenamefont {Ciliberto}, \citenamefont {Baudet}, \citenamefont {Chavarria},\ and\ \citenamefont {Tripiccione}}]{benzi1993extended2}%
  \BibitemOpen
  \bibfield  {author} {\bibinfo {author} {\bibfnamefont {R.}~\bibnamefont {Benzi}}, \bibinfo {author} {\bibfnamefont {S.}~\bibnamefont {Ciliberto}}, \bibinfo {author} {\bibfnamefont {C.}~\bibnamefont {Baudet}}, \bibinfo {author} {\bibfnamefont {G.~R.}\ \bibnamefont {Chavarria}},\ and\ \bibinfo {author} {\bibfnamefont {R.}~\bibnamefont {Tripiccione}},\ }\href@noop {} {\bibfield  {journal} {\bibinfo  {journal} {Europhysics letters}\ }\textbf {\bibinfo {volume} {24}},\ \bibinfo {pages} {275} (\bibinfo {year} {1993}{\natexlab{b}})}\BibitemShut {NoStop}%
\bibitem [{\citenamefont {Benzi}\ \emph {et~al.}(1991)\citenamefont {Benzi}, \citenamefont {Biferale}, \citenamefont {Paladin}, \citenamefont {Vulpiani},\ and\ \citenamefont {Vergassola}}]{benzi1991multifractality}%
  \BibitemOpen
  \bibfield  {author} {\bibinfo {author} {\bibfnamefont {R.}~\bibnamefont {Benzi}}, \bibinfo {author} {\bibfnamefont {L.}~\bibnamefont {Biferale}}, \bibinfo {author} {\bibfnamefont {G.}~\bibnamefont {Paladin}}, \bibinfo {author} {\bibfnamefont {A.}~\bibnamefont {Vulpiani}},\ and\ \bibinfo {author} {\bibfnamefont {M.}~\bibnamefont {Vergassola}},\ }\href@noop {} {\bibfield  {journal} {\bibinfo  {journal} {Physical review letters}\ }\textbf {\bibinfo {volume} {67}},\ \bibinfo {pages} {2299} (\bibinfo {year} {1991})}\BibitemShut {NoStop}%
\bibitem [{\citenamefont {Nikora}\ and\ \citenamefont {Goring}(2001)}]{nikora2001extended}%
  \BibitemOpen
  \bibfield  {author} {\bibinfo {author} {\bibfnamefont {V.~I.}\ \bibnamefont {Nikora}}\ and\ \bibinfo {author} {\bibfnamefont {D.~G.}\ \bibnamefont {Goring}},\ }\href@noop {} {\bibfield  {journal} {\bibinfo  {journal} {Mathematical geology}\ }\textbf {\bibinfo {volume} {33}},\ \bibinfo {pages} {251} (\bibinfo {year} {2001})}\BibitemShut {NoStop}%
\bibitem [{\citenamefont {Carbone}\ \emph {et~al.}(1996)\citenamefont {Carbone}, \citenamefont {Bruno},\ and\ \citenamefont {Veltri}}]{carbone1996evidences}%
  \BibitemOpen
  \bibfield  {author} {\bibinfo {author} {\bibfnamefont {V.}~\bibnamefont {Carbone}}, \bibinfo {author} {\bibfnamefont {R.}~\bibnamefont {Bruno}},\ and\ \bibinfo {author} {\bibfnamefont {P.}~\bibnamefont {Veltri}},\ }\href@noop {} {\bibfield  {journal} {\bibinfo  {journal} {Geophysical research letters}\ }\textbf {\bibinfo {volume} {23}},\ \bibinfo {pages} {121} (\bibinfo {year} {1996})}\BibitemShut {NoStop}%
\bibitem [{\citenamefont {Kiliyanpilakkil}\ and\ \citenamefont {Basu}(2016)}]{kiliyanpilakkil2016extended}%
  \BibitemOpen
  \bibfield  {author} {\bibinfo {author} {\bibfnamefont {V.}~\bibnamefont {Kiliyanpilakkil}}\ and\ \bibinfo {author} {\bibfnamefont {S.}~\bibnamefont {Basu}},\ }\href@noop {} {\bibfield  {journal} {\bibinfo  {journal} {Europhysics Letters}\ }\textbf {\bibinfo {volume} {112}},\ \bibinfo {pages} {64003} (\bibinfo {year} {2016})}\BibitemShut {NoStop}%
\bibitem [{\citenamefont {Zhao}\ \emph {et~al.}(2025{\natexlab{a}})\citenamefont {Zhao}, \citenamefont {Zeng},\ and\ \citenamefont {Lin}}]{zhao2025turbulence}%
  \BibitemOpen
  \bibfield  {author} {\bibinfo {author} {\bibfnamefont {M.}~\bibnamefont {Zhao}}, \bibinfo {author} {\bibfnamefont {X.}~\bibnamefont {Zeng}},\ and\ \bibinfo {author} {\bibfnamefont {Y.}~\bibnamefont {Lin}},\ }\href@noop {} {\bibfield  {journal} {\bibinfo  {journal} {arXiv preprint arXiv:2508.06143}\ } (\bibinfo {year} {2025}{\natexlab{a}})}\BibitemShut {NoStop}%
\bibitem [{\citenamefont {Zhao}(2025)}]{zhao2025spatiotemporal}%
  \BibitemOpen
  \bibfield  {author} {\bibinfo {author} {\bibfnamefont {M.}~\bibnamefont {Zhao}},\ }\href@noop {} {\bibfield  {journal} {\bibinfo  {journal} {Physical Review A}\ }\textbf {\bibinfo {volume} {111}},\ \bibinfo {pages} {033320} (\bibinfo {year} {2025})}\BibitemShut {NoStop}%
\bibitem [{\citenamefont {Zhao}\ \emph {et~al.}(2025{\natexlab{b}})\citenamefont {Zhao}, \citenamefont {Tao},\ and\ \citenamefont {Spielman}}]{zhao2025kolmogorov}%
  \BibitemOpen
  \bibfield  {author} {\bibinfo {author} {\bibfnamefont {M.}~\bibnamefont {Zhao}}, \bibinfo {author} {\bibfnamefont {J.}~\bibnamefont {Tao}},\ and\ \bibinfo {author} {\bibfnamefont {I.}~\bibnamefont {Spielman}},\ }\href@noop {} {\bibfield  {journal} {\bibinfo  {journal} {Physical Review Letters}\ }\textbf {\bibinfo {volume} {134}},\ \bibinfo {pages} {083402} (\bibinfo {year} {2025}{\natexlab{b}})}\BibitemShut {NoStop}%
\bibitem [{\citenamefont {Varga}\ \emph {et~al.}(2018)\citenamefont {Varga}, \citenamefont {Gao}, \citenamefont {Guo},\ and\ \citenamefont {Skrbek}}]{varga2018intermittency}%
  \BibitemOpen
  \bibfield  {author} {\bibinfo {author} {\bibfnamefont {E.}~\bibnamefont {Varga}}, \bibinfo {author} {\bibfnamefont {J.}~\bibnamefont {Gao}}, \bibinfo {author} {\bibfnamefont {W.}~\bibnamefont {Guo}},\ and\ \bibinfo {author} {\bibfnamefont {L.}~\bibnamefont {Skrbek}},\ }\href@noop {} {\bibfield  {journal} {\bibinfo  {journal} {Physical Review Fluids}\ }\textbf {\bibinfo {volume} {3}},\ \bibinfo {pages} {094601} (\bibinfo {year} {2018})}\BibitemShut {NoStop}%
\bibitem [{\citenamefont {Budaev}\ \emph {et~al.}(2008)\citenamefont {Budaev}, \citenamefont {Ohno}, \citenamefont {Masuzaki}, \citenamefont {Morisaki}, \citenamefont {Komori},\ and\ \citenamefont {Takamura}}]{budaev2008extended}%
  \BibitemOpen
  \bibfield  {author} {\bibinfo {author} {\bibfnamefont {V.}~\bibnamefont {Budaev}}, \bibinfo {author} {\bibfnamefont {N.}~\bibnamefont {Ohno}}, \bibinfo {author} {\bibfnamefont {S.}~\bibnamefont {Masuzaki}}, \bibinfo {author} {\bibfnamefont {T.}~\bibnamefont {Morisaki}}, \bibinfo {author} {\bibfnamefont {A.}~\bibnamefont {Komori}},\ and\ \bibinfo {author} {\bibfnamefont {S.}~\bibnamefont {Takamura}},\ }\href@noop {} {\bibfield  {journal} {\bibinfo  {journal} {Nuclear Fusion}\ }\textbf {\bibinfo {volume} {48}},\ \bibinfo {pages} {024014} (\bibinfo {year} {2008})}\BibitemShut {NoStop}%
\bibitem [{\citenamefont {Constantin}\ and\ \citenamefont {Das~Sarma}(2005)}]{constantin2005volatility}%
  \BibitemOpen
  \bibfield  {author} {\bibinfo {author} {\bibfnamefont {M.}~\bibnamefont {Constantin}}\ and\ \bibinfo {author} {\bibfnamefont {S.}~\bibnamefont {Das~Sarma}},\ }\href@noop {} {\bibfield  {journal} {\bibinfo  {journal} {Physical Review E—Statistical, Nonlinear, and Soft Matter Physics}\ }\textbf {\bibinfo {volume} {72}},\ \bibinfo {pages} {051106} (\bibinfo {year} {2005})}\BibitemShut {NoStop}%
\bibitem [{\citenamefont {Kolmogorov}(1941)}]{kolmogorov1941local}%
  \BibitemOpen
  \bibfield  {author} {\bibinfo {author} {\bibfnamefont {A.~N.}\ \bibnamefont {Kolmogorov}},\ }\href@noop {} {\bibfield  {journal} {\bibinfo  {journal} {Numbers. In Dokl. Akad. Nauk SSSR}\ }\textbf {\bibinfo {volume} {30}},\ \bibinfo {pages} {301} (\bibinfo {year} {1941})}\BibitemShut {NoStop}%
\bibitem [{\citenamefont {Yakhot}(2001)}]{yakhot2001mean}%
  \BibitemOpen
  \bibfield  {author} {\bibinfo {author} {\bibfnamefont {V.}~\bibnamefont {Yakhot}},\ }\href@noop {} {\bibfield  {journal} {\bibinfo  {journal} {Physical Review Letters}\ }\textbf {\bibinfo {volume} {87}},\ \bibinfo {pages} {234501} (\bibinfo {year} {2001})}\BibitemShut {NoStop}%
\bibitem [{\citenamefont {Ching}\ \emph {et~al.}(2002)\citenamefont {Ching}, \citenamefont {She}, \citenamefont {Su},\ and\ \citenamefont {Zou}}]{ching2002extended}%
  \BibitemOpen
  \bibfield  {author} {\bibinfo {author} {\bibfnamefont {E.~S.}\ \bibnamefont {Ching}}, \bibinfo {author} {\bibfnamefont {Z.-S.}\ \bibnamefont {She}}, \bibinfo {author} {\bibfnamefont {W.}~\bibnamefont {Su}},\ and\ \bibinfo {author} {\bibfnamefont {Z.}~\bibnamefont {Zou}},\ }\href@noop {} {\bibfield  {journal} {\bibinfo  {journal} {Physical Review E}\ }\textbf {\bibinfo {volume} {65}},\ \bibinfo {pages} {066303} (\bibinfo {year} {2002})}\BibitemShut {NoStop}%
\bibitem [{\citenamefont {Chakraborty}\ \emph {et~al.}(2010)\citenamefont {Chakraborty}, \citenamefont {Frisch},\ and\ \citenamefont {Ray}}]{chakraborty2010extended}%
  \BibitemOpen
  \bibfield  {author} {\bibinfo {author} {\bibfnamefont {S.}~\bibnamefont {Chakraborty}}, \bibinfo {author} {\bibfnamefont {U.}~\bibnamefont {Frisch}},\ and\ \bibinfo {author} {\bibfnamefont {S.~S.}\ \bibnamefont {Ray}},\ }\href@noop {} {\bibfield  {journal} {\bibinfo  {journal} {Journal of fluid mechanics}\ }\textbf {\bibinfo {volume} {649}},\ \bibinfo {pages} {275} (\bibinfo {year} {2010})}\BibitemShut {NoStop}%
\bibitem [{\citenamefont {Hill}\ \emph {et~al.}(1980)\citenamefont {Hill}, \citenamefont {Tremblay},\ and\ \citenamefont {Kawasaki}}]{hill1980modal}%
  \BibitemOpen
  \bibfield  {author} {\bibinfo {author} {\bibfnamefont {K.~O.}\ \bibnamefont {Hill}}, \bibinfo {author} {\bibfnamefont {Y.}~\bibnamefont {Tremblay}},\ and\ \bibinfo {author} {\bibfnamefont {B.~S.}\ \bibnamefont {Kawasaki}},\ }\href@noop {} {\bibfield  {journal} {\bibinfo  {journal} {Optics letters}\ }\textbf {\bibinfo {volume} {5}},\ \bibinfo {pages} {270} (\bibinfo {year} {1980})}\BibitemShut {NoStop}%
\bibitem [{\citenamefont {Daino}\ \emph {et~al.}(1980)\citenamefont {Daino}, \citenamefont {De~Marchis},\ and\ \citenamefont {Piazzolla}}]{daino1980speckle}%
  \BibitemOpen
  \bibfield  {author} {\bibinfo {author} {\bibfnamefont {B.}~\bibnamefont {Daino}}, \bibinfo {author} {\bibfnamefont {G.}~\bibnamefont {De~Marchis}},\ and\ \bibinfo {author} {\bibfnamefont {S.}~\bibnamefont {Piazzolla}},\ }\href@noop {} {\bibfield  {journal} {\bibinfo  {journal} {Optica Acta: International Journal of Optics}\ }\textbf {\bibinfo {volume} {27}},\ \bibinfo {pages} {1151} (\bibinfo {year} {1980})}\BibitemShut {NoStop}%
\bibitem [{\citenamefont {Rawson}\ \emph {et~al.}(1980)\citenamefont {Rawson}, \citenamefont {Goodman},\ and\ \citenamefont {Norton}}]{rawson1980frequency}%
  \BibitemOpen
  \bibfield  {author} {\bibinfo {author} {\bibfnamefont {E.~G.}\ \bibnamefont {Rawson}}, \bibinfo {author} {\bibfnamefont {J.~W.}\ \bibnamefont {Goodman}},\ and\ \bibinfo {author} {\bibfnamefont {R.~E.}\ \bibnamefont {Norton}},\ }\href@noop {} {\bibfield  {journal} {\bibinfo  {journal} {Journal of the Optical Society of America}\ }\textbf {\bibinfo {volume} {70}},\ \bibinfo {pages} {968} (\bibinfo {year} {1980})}\BibitemShut {NoStop}%
\bibitem [{\citenamefont {Tremblay}\ \emph {et~al.}(1981)\citenamefont {Tremblay}, \citenamefont {Kawasaki},\ and\ \citenamefont {Hill}}]{tremblay1981modal}%
  \BibitemOpen
  \bibfield  {author} {\bibinfo {author} {\bibfnamefont {Y.}~\bibnamefont {Tremblay}}, \bibinfo {author} {\bibfnamefont {B.}~\bibnamefont {Kawasaki}},\ and\ \bibinfo {author} {\bibfnamefont {K.}~\bibnamefont {Hill}},\ }\href@noop {} {\bibfield  {journal} {\bibinfo  {journal} {Applied optics}\ }\textbf {\bibinfo {volume} {20}},\ \bibinfo {pages} {1652} (\bibinfo {year} {1981})}\BibitemShut {NoStop}%
\bibitem [{\citenamefont {Takai}\ and\ \citenamefont {Asakura}(1985)}]{takai1985statistical}%
  \BibitemOpen
  \bibfield  {author} {\bibinfo {author} {\bibfnamefont {N.}~\bibnamefont {Takai}}\ and\ \bibinfo {author} {\bibfnamefont {T.}~\bibnamefont {Asakura}},\ }\href@noop {} {\bibfield  {journal} {\bibinfo  {journal} {Journal of the Optical Society of America A}\ }\textbf {\bibinfo {volume} {2}},\ \bibinfo {pages} {1282} (\bibinfo {year} {1985})}\BibitemShut {NoStop}%
\bibitem [{\citenamefont {Borhani}\ \emph {et~al.}(2018)\citenamefont {Borhani}, \citenamefont {Kakkava}, \citenamefont {Moser},\ and\ \citenamefont {Psaltis}}]{borhani2018learning}%
  \BibitemOpen
  \bibfield  {author} {\bibinfo {author} {\bibfnamefont {N.}~\bibnamefont {Borhani}}, \bibinfo {author} {\bibfnamefont {E.}~\bibnamefont {Kakkava}}, \bibinfo {author} {\bibfnamefont {C.}~\bibnamefont {Moser}},\ and\ \bibinfo {author} {\bibfnamefont {D.}~\bibnamefont {Psaltis}},\ }\href@noop {} {\bibfield  {journal} {\bibinfo  {journal} {Optica}\ }\textbf {\bibinfo {volume} {5}},\ \bibinfo {pages} {960} (\bibinfo {year} {2018})}\BibitemShut {NoStop}%
\bibitem [{\citenamefont {Rahmani}\ \emph {et~al.}(2018)\citenamefont {Rahmani}, \citenamefont {Loterie}, \citenamefont {Konstantinou}, \citenamefont {Psaltis},\ and\ \citenamefont {Moser}}]{rahmani2018multimode}%
  \BibitemOpen
  \bibfield  {author} {\bibinfo {author} {\bibfnamefont {B.}~\bibnamefont {Rahmani}}, \bibinfo {author} {\bibfnamefont {D.}~\bibnamefont {Loterie}}, \bibinfo {author} {\bibfnamefont {G.}~\bibnamefont {Konstantinou}}, \bibinfo {author} {\bibfnamefont {D.}~\bibnamefont {Psaltis}},\ and\ \bibinfo {author} {\bibfnamefont {C.}~\bibnamefont {Moser}},\ }\href@noop {} {\bibfield  {journal} {\bibinfo  {journal} {Light: science \& applications}\ }\textbf {\bibinfo {volume} {7}},\ \bibinfo {pages} {69} (\bibinfo {year} {2018})}\BibitemShut {NoStop}%
\bibitem [{\citenamefont {Moran}\ \emph {et~al.}(2018)\citenamefont {Moran}, \citenamefont {Caramazza}, \citenamefont {Faccio},\ and\ \citenamefont {Murray-Smith}}]{moran2018deep}%
  \BibitemOpen
  \bibfield  {author} {\bibinfo {author} {\bibfnamefont {O.}~\bibnamefont {Moran}}, \bibinfo {author} {\bibfnamefont {P.}~\bibnamefont {Caramazza}}, \bibinfo {author} {\bibfnamefont {D.}~\bibnamefont {Faccio}},\ and\ \bibinfo {author} {\bibfnamefont {R.}~\bibnamefont {Murray-Smith}},\ }\href@noop {} {\bibfield  {journal} {\bibinfo  {journal} {Advances in neural information processing systems}\ }\textbf {\bibinfo {volume} {31}} (\bibinfo {year} {2018})}\BibitemShut {NoStop}%
\bibitem [{\citenamefont {Goodman}(1975)}]{goodman1975statistical}%
  \BibitemOpen
  \bibfield  {author} {\bibinfo {author} {\bibfnamefont {J.~W.}\ \bibnamefont {Goodman}},\ }in\ \href@noop {} {\emph {\bibinfo {booktitle} {Laser speckle and related phenomena}}}\ (\bibinfo  {publisher} {Springer},\ \bibinfo {year} {1975})\ pp.\ \bibinfo {pages} {9--75}\BibitemShut {NoStop}%
\bibitem [{\citenamefont {Tsuji}\ \emph {et~al.}(1984)\citenamefont {Tsuji}, \citenamefont {Asakura},\ and\ \citenamefont {Fujii}}]{tsuji1984variation}%
  \BibitemOpen
  \bibfield  {author} {\bibinfo {author} {\bibfnamefont {T.}~\bibnamefont {Tsuji}}, \bibinfo {author} {\bibfnamefont {T.}~\bibnamefont {Asakura}},\ and\ \bibinfo {author} {\bibfnamefont {H.}~\bibnamefont {Fujii}},\ }\href@noop {} {\bibfield  {journal} {\bibinfo  {journal} {Optical and quantum electronics}\ }\textbf {\bibinfo {volume} {16}},\ \bibinfo {pages} {197} (\bibinfo {year} {1984})}\BibitemShut {NoStop}%
\bibitem [{\citenamefont {Imai}(1986)}]{imai1986statistical}%
  \BibitemOpen
  \bibfield  {author} {\bibinfo {author} {\bibfnamefont {M.}~\bibnamefont {Imai}},\ }\href@noop {} {\bibfield  {journal} {\bibinfo  {journal} {Research Reports of the Faculty of Engineering, Hokkaido University}\ }\textbf {\bibinfo {volume} {130}},\ \bibinfo {pages} {89} (\bibinfo {year} {1986})}\BibitemShut {NoStop}%
\bibitem [{\citenamefont {Goodman}(2007)}]{goodman2007speckle}%
  \BibitemOpen
  \bibfield  {author} {\bibinfo {author} {\bibfnamefont {J.~W.}\ \bibnamefont {Goodman}},\ }\href@noop {} {\emph {\bibinfo {title} {Speckle phenomena in optics: theory and applications}}}\ (\bibinfo  {publisher} {Roberts and Company Publishers},\ \bibinfo {year} {2007})\BibitemShut {NoStop}%
\bibitem [{\citenamefont {Regez}\ \emph {et~al.}(2009)\citenamefont {Regez}, \citenamefont {Sayeh}, \citenamefont {Mahajan},\ and\ \citenamefont {Figueroa}}]{regez2009novel}%
  \BibitemOpen
  \bibfield  {author} {\bibinfo {author} {\bibfnamefont {B.}~\bibnamefont {Regez}}, \bibinfo {author} {\bibfnamefont {M.}~\bibnamefont {Sayeh}}, \bibinfo {author} {\bibfnamefont {A.}~\bibnamefont {Mahajan}},\ and\ \bibinfo {author} {\bibfnamefont {F.}~\bibnamefont {Figueroa}},\ }\href@noop {} {\bibfield  {journal} {\bibinfo  {journal} {Measurement}\ }\textbf {\bibinfo {volume} {42}},\ \bibinfo {pages} {183} (\bibinfo {year} {2009})}\BibitemShut {NoStop}%
\bibitem [{\citenamefont {Fujiwara}\ \emph {et~al.}(2018)\citenamefont {Fujiwara}, \citenamefont {Evaristo~da Silva}, \citenamefont {R.~Marques},\ and\ \citenamefont {B.~Cordeiro}}]{fujiwara2018polymer}%
  \BibitemOpen
  \bibfield  {author} {\bibinfo {author} {\bibfnamefont {E.}~\bibnamefont {Fujiwara}}, \bibinfo {author} {\bibfnamefont {L.}~\bibnamefont {Evaristo~da Silva}}, \bibinfo {author} {\bibfnamefont {T.~H.}\ \bibnamefont {R.~Marques}},\ and\ \bibinfo {author} {\bibfnamefont {C.~M.}\ \bibnamefont {B.~Cordeiro}},\ }\href@noop {} {\bibfield  {journal} {\bibinfo  {journal} {Optical Engineering}\ }\textbf {\bibinfo {volume} {57}},\ \bibinfo {pages} {116107} (\bibinfo {year} {2018})}\BibitemShut {NoStop}%
\bibitem [{\citenamefont {Musin}\ \emph {et~al.}(2016)\citenamefont {Musin}, \citenamefont {M{\'e}gret},\ and\ \citenamefont {Wuilpart}}]{musin2016fiber}%
  \BibitemOpen
  \bibfield  {author} {\bibinfo {author} {\bibfnamefont {F.}~\bibnamefont {Musin}}, \bibinfo {author} {\bibfnamefont {P.}~\bibnamefont {M{\'e}gret}},\ and\ \bibinfo {author} {\bibfnamefont {M.}~\bibnamefont {Wuilpart}},\ }\href@noop {} {\bibfield  {journal} {\bibinfo  {journal} {Sensors}\ }\textbf {\bibinfo {volume} {16}},\ \bibinfo {pages} {1189} (\bibinfo {year} {2016})}\BibitemShut {NoStop}%
\bibitem [{\citenamefont {Redding}\ \emph {et~al.}(2013)\citenamefont {Redding}, \citenamefont {Popoff},\ and\ \citenamefont {Cao}}]{redding2013all}%
  \BibitemOpen
  \bibfield  {author} {\bibinfo {author} {\bibfnamefont {B.}~\bibnamefont {Redding}}, \bibinfo {author} {\bibfnamefont {S.~M.}\ \bibnamefont {Popoff}},\ and\ \bibinfo {author} {\bibfnamefont {H.}~\bibnamefont {Cao}},\ }\href@noop {} {\bibfield  {journal} {\bibinfo  {journal} {Optics express}\ }\textbf {\bibinfo {volume} {21}},\ \bibinfo {pages} {6584} (\bibinfo {year} {2013})}\BibitemShut {NoStop}%
\bibitem [{\citenamefont {Wang}\ \emph {et~al.}(2023)\citenamefont {Wang}, \citenamefont {Mizuno}, \citenamefont {Dong}, \citenamefont {Kurz}, \citenamefont {K{\"o}hler}, \citenamefont {Kienle}, \citenamefont {Lee}, \citenamefont {Jakobi},\ and\ \citenamefont {Koch}}]{wang2023multimode}%
  \BibitemOpen
  \bibfield  {author} {\bibinfo {author} {\bibfnamefont {K.}~\bibnamefont {Wang}}, \bibinfo {author} {\bibfnamefont {Y.}~\bibnamefont {Mizuno}}, \bibinfo {author} {\bibfnamefont {X.}~\bibnamefont {Dong}}, \bibinfo {author} {\bibfnamefont {W.}~\bibnamefont {Kurz}}, \bibinfo {author} {\bibfnamefont {M.}~\bibnamefont {K{\"o}hler}}, \bibinfo {author} {\bibfnamefont {P.}~\bibnamefont {Kienle}}, \bibinfo {author} {\bibfnamefont {H.}~\bibnamefont {Lee}}, \bibinfo {author} {\bibfnamefont {M.}~\bibnamefont {Jakobi}},\ and\ \bibinfo {author} {\bibfnamefont {A.~W.}\ \bibnamefont {Koch}},\ }\href@noop {} {\bibfield  {journal} {\bibinfo  {journal} {Measurement Science and Technology}\ }\textbf {\bibinfo {volume} {35}},\ \bibinfo {pages} {022002} (\bibinfo {year} {2023})}\BibitemShut {NoStop}%
\bibitem [{\citenamefont {Newaz}\ \emph {et~al.}(2023)\citenamefont {Newaz}, \citenamefont {Faruque}, \citenamefont {Al~Mahmud}, \citenamefont {Sagor},\ and\ \citenamefont {Khan}}]{newaz2023machine}%
  \BibitemOpen
  \bibfield  {author} {\bibinfo {author} {\bibfnamefont {A.}~\bibnamefont {Newaz}}, \bibinfo {author} {\bibfnamefont {M.~O.}\ \bibnamefont {Faruque}}, \bibinfo {author} {\bibfnamefont {R.}~\bibnamefont {Al~Mahmud}}, \bibinfo {author} {\bibfnamefont {R.~H.}\ \bibnamefont {Sagor}},\ and\ \bibinfo {author} {\bibfnamefont {M.~Z.~M.}\ \bibnamefont {Khan}},\ }\href@noop {} {\bibfield  {journal} {\bibinfo  {journal} {IEEE Sensors Journal}\ }\textbf {\bibinfo {volume} {23}},\ \bibinfo {pages} {20937} (\bibinfo {year} {2023})}\BibitemShut {NoStop}%
\bibitem [{\citenamefont {Popoff}\ \emph {et~al.}(2010)\citenamefont {Popoff}, \citenamefont {Lerosey}, \citenamefont {Carminati}, \citenamefont {Fink}, \citenamefont {Boccara},\ and\ \citenamefont {Gigan}}]{popoff2010measuring}%
  \BibitemOpen
  \bibfield  {author} {\bibinfo {author} {\bibfnamefont {S.~M.}\ \bibnamefont {Popoff}}, \bibinfo {author} {\bibfnamefont {G.}~\bibnamefont {Lerosey}}, \bibinfo {author} {\bibfnamefont {R.}~\bibnamefont {Carminati}}, \bibinfo {author} {\bibfnamefont {M.}~\bibnamefont {Fink}}, \bibinfo {author} {\bibfnamefont {A.~C.}\ \bibnamefont {Boccara}},\ and\ \bibinfo {author} {\bibfnamefont {S.}~\bibnamefont {Gigan}},\ }\href@noop {} {\bibfield  {journal} {\bibinfo  {journal} {Physical review letters}\ }\textbf {\bibinfo {volume} {104}},\ \bibinfo {pages} {100601} (\bibinfo {year} {2010})}\BibitemShut {NoStop}%
\bibitem [{\citenamefont {Carpenter}\ \emph {et~al.}(2015)\citenamefont {Carpenter}, \citenamefont {Eggleton},\ and\ \citenamefont {Schr{\"o}der}}]{carpenter2015observation}%
  \BibitemOpen
  \bibfield  {author} {\bibinfo {author} {\bibfnamefont {J.}~\bibnamefont {Carpenter}}, \bibinfo {author} {\bibfnamefont {B.~J.}\ \bibnamefont {Eggleton}},\ and\ \bibinfo {author} {\bibfnamefont {J.}~\bibnamefont {Schr{\"o}der}},\ }\href@noop {} {\bibfield  {journal} {\bibinfo  {journal} {Nature Photonics}\ }\textbf {\bibinfo {volume} {9}},\ \bibinfo {pages} {751} (\bibinfo {year} {2015})}\BibitemShut {NoStop}%
\bibitem [{\citenamefont {Aluie}(2011)}]{aluie2011compressible}%
  \BibitemOpen
  \bibfield  {author} {\bibinfo {author} {\bibfnamefont {H.}~\bibnamefont {Aluie}},\ }\href@noop {} {\bibfield  {journal} {\bibinfo  {journal} {Physical review letters}\ }\textbf {\bibinfo {volume} {106}},\ \bibinfo {pages} {174502} (\bibinfo {year} {2011})}\BibitemShut {NoStop}%
\bibitem [{\citenamefont {Kolmogorov}(1962)}]{kolmogorov1962refinement}%
  \BibitemOpen
  \bibfield  {author} {\bibinfo {author} {\bibfnamefont {A.~N.}\ \bibnamefont {Kolmogorov}},\ }\href@noop {} {\bibfield  {journal} {\bibinfo  {journal} {Journal of Fluid Mechanics}\ }\textbf {\bibinfo {volume} {13}},\ \bibinfo {pages} {82} (\bibinfo {year} {1962})}\BibitemShut {NoStop}%
\bibitem [{\citenamefont {Obukhov}(1962)}]{obukhov1962some}%
  \BibitemOpen
  \bibfield  {author} {\bibinfo {author} {\bibfnamefont {A.}~\bibnamefont {Obukhov}},\ }\href@noop {} {\bibfield  {journal} {\bibinfo  {journal} {Journal of Geophysical Research}\ }\textbf {\bibinfo {volume} {67}},\ \bibinfo {pages} {3011} (\bibinfo {year} {1962})}\BibitemShut {NoStop}%
\bibitem [{\citenamefont {Gloge}(1971)}]{gloge1971weakly}%
  \BibitemOpen
  \bibfield  {author} {\bibinfo {author} {\bibfnamefont {D.}~\bibnamefont {Gloge}},\ }\href@noop {} {\bibfield  {journal} {\bibinfo  {journal} {Applied optics}\ }\textbf {\bibinfo {volume} {10}},\ \bibinfo {pages} {2252} (\bibinfo {year} {1971})}\BibitemShut {NoStop}%
\bibitem [{\citenamefont {Ghatak}\ and\ \citenamefont {Thyagarajan}(1998)}]{ghatak1998introduction}%
  \BibitemOpen
  \bibfield  {author} {\bibinfo {author} {\bibfnamefont {A.~K.}\ \bibnamefont {Ghatak}}\ and\ \bibinfo {author} {\bibfnamefont {K.}~\bibnamefont {Thyagarajan}},\ }\href@noop {} {\emph {\bibinfo {title} {An introduction to fiber optics}}}\ (\bibinfo  {publisher} {Cambridge university press},\ \bibinfo {year} {1998})\BibitemShut {NoStop}%
\bibitem [{\citenamefont {Snyder}(1972)}]{snyder1972coupled}%
  \BibitemOpen
  \bibfield  {author} {\bibinfo {author} {\bibfnamefont {A.~W.}\ \bibnamefont {Snyder}},\ }\href@noop {} {\bibfield  {journal} {\bibinfo  {journal} {Journal of the optical society of America}\ }\textbf {\bibinfo {volume} {62}},\ \bibinfo {pages} {1267} (\bibinfo {year} {1972})}\BibitemShut {NoStop}%
\bibitem [{\citenamefont {Dormand}\ and\ \citenamefont {Prince}(1980)}]{dormand1980family}%
  \BibitemOpen
  \bibfield  {author} {\bibinfo {author} {\bibfnamefont {J.~R.}\ \bibnamefont {Dormand}}\ and\ \bibinfo {author} {\bibfnamefont {P.~J.}\ \bibnamefont {Prince}},\ }\href@noop {} {\bibfield  {journal} {\bibinfo  {journal} {Journal of computational and applied mathematics}\ }\textbf {\bibinfo {volume} {6}},\ \bibinfo {pages} {19} (\bibinfo {year} {1980})}\BibitemShut {NoStop}%
\bibitem [{\citenamefont {Shampine}(1986)}]{shampine1986some}%
  \BibitemOpen
  \bibfield  {author} {\bibinfo {author} {\bibfnamefont {L.~F.}\ \bibnamefont {Shampine}},\ }\href@noop {} {\bibfield  {journal} {\bibinfo  {journal} {Mathematics of computation}\ }\textbf {\bibinfo {volume} {46}},\ \bibinfo {pages} {135} (\bibinfo {year} {1986})}\BibitemShut {NoStop}%
\bibitem [{\citenamefont {Bengtsson}\ and\ \citenamefont {{\.Z}yczkowski}(2017)}]{bengtsson2017geometry}%
  \BibitemOpen
  \bibfield  {author} {\bibinfo {author} {\bibfnamefont {I.}~\bibnamefont {Bengtsson}}\ and\ \bibinfo {author} {\bibfnamefont {K.}~\bibnamefont {{\.Z}yczkowski}},\ }\href@noop {} {\emph {\bibinfo {title} {Geometry of quantum states: an introduction to quantum entanglement}}}\ (\bibinfo  {publisher} {Cambridge university press},\ \bibinfo {year} {2017})\BibitemShut {NoStop}%
\bibitem [{\citenamefont {Krupa}\ \emph {et~al.}(2017)\citenamefont {Krupa}, \citenamefont {Tonello}, \citenamefont {Shalaby}, \citenamefont {Fabert}, \citenamefont {Barth{\'e}l{\'e}my}, \citenamefont {Millot}, \citenamefont {Wabnitz},\ and\ \citenamefont {Couderc}}]{krupa2017spatial}%
  \BibitemOpen
  \bibfield  {author} {\bibinfo {author} {\bibfnamefont {K.}~\bibnamefont {Krupa}}, \bibinfo {author} {\bibfnamefont {A.}~\bibnamefont {Tonello}}, \bibinfo {author} {\bibfnamefont {B.~M.}\ \bibnamefont {Shalaby}}, \bibinfo {author} {\bibfnamefont {M.}~\bibnamefont {Fabert}}, \bibinfo {author} {\bibfnamefont {A.}~\bibnamefont {Barth{\'e}l{\'e}my}}, \bibinfo {author} {\bibfnamefont {G.}~\bibnamefont {Millot}}, \bibinfo {author} {\bibfnamefont {S.}~\bibnamefont {Wabnitz}},\ and\ \bibinfo {author} {\bibfnamefont {V.}~\bibnamefont {Couderc}},\ }\href@noop {} {\bibfield  {journal} {\bibinfo  {journal} {Nature photonics}\ }\textbf {\bibinfo {volume} {11}},\ \bibinfo {pages} {237} (\bibinfo {year} {2017})}\BibitemShut {NoStop}%
\bibitem [{\citenamefont {Podivilov}\ \emph {et~al.}(2019)\citenamefont {Podivilov}, \citenamefont {Kharenko}, \citenamefont {Gonta}, \citenamefont {Krupa}, \citenamefont {Sidelnikov}, \citenamefont {Turitsyn}, \citenamefont {Fedoruk}, \citenamefont {Babin},\ and\ \citenamefont {Wabnitz}}]{PhysRevLett.122.103902}%
  \BibitemOpen
  \bibfield  {author} {\bibinfo {author} {\bibfnamefont {E.~V.}\ \bibnamefont {Podivilov}}, \bibinfo {author} {\bibfnamefont {D.~S.}\ \bibnamefont {Kharenko}}, \bibinfo {author} {\bibfnamefont {V.~A.}\ \bibnamefont {Gonta}}, \bibinfo {author} {\bibfnamefont {K.}~\bibnamefont {Krupa}}, \bibinfo {author} {\bibfnamefont {O.~S.}\ \bibnamefont {Sidelnikov}}, \bibinfo {author} {\bibfnamefont {S.}~\bibnamefont {Turitsyn}}, \bibinfo {author} {\bibfnamefont {M.~P.}\ \bibnamefont {Fedoruk}}, \bibinfo {author} {\bibfnamefont {S.~A.}\ \bibnamefont {Babin}},\ and\ \bibinfo {author} {\bibfnamefont {S.}~\bibnamefont {Wabnitz}},\ }\href {https://doi.org/10.1103/PhysRevLett.122.103902} {\bibfield  {journal} {\bibinfo  {journal} {Phys. Rev. Lett.}\ }\textbf {\bibinfo {volume} {122}},\ \bibinfo {pages} {103902} (\bibinfo {year} {2019})}\BibitemShut {NoStop}%
\end{thebibliography}
